\documentclass{aastex}
\usepackage{graphicx}
\usepackage{times}
\begin{document}


\title{
A Search for Eclipsing Binary Lightcurve Variations among MACHO Project
Lightcurves of
3256 Fundamental-Mode RR Lyrae Variables in the Galactic Bulge
}

\author{Michael W. Richmond}
\shorttitle{A search for eclipsing RR Lyraes}
\shortauthors{M. W. Richmond}
\affil{
Physics Department, Rochester Institute of Technology,
84 Lomb Memorial Drive, Rochester, NY, 14623 USA
}
\email{mwrsps@rit.edu}


\begin{abstract}
The MACHO Project collected photometry of many RR Lyrae stars
from its observations of the Milky Way's bulge.
We examined the lightcurves of 3256 stars identified as RRab Lyr 
variables by 
\cite{Kund2008},
subtracting an empirical model of the pulsation lightcurve and
searching for periodic variation in the residuals.
There are no systems which show the brief dips in 
light characteristic of detached eclipsing binary systems.
We discuss the results for objects which show the largest
residual periodic modulation,
most of which are probably due to aliases of the 
fundamental period.
\end{abstract}


\section{Introduction}
Measuring the distances to objects is one of the most
fundamental tasks in astronomy,
yet it is also one of the most difficult.
Among the many indirect methods astronomers have devised
for dealing with this problem is the technique 
of ``standard candles:''
identifying a class of sources which have the same luminosity
and are easy to recognize.
RR Lyrae stars fit into this category:
they vary in brightness 
by a considerable amount  
(amplitudes of order half a magnitude)
in a short time (periods of order half a day).
Moreover, their light curves exhibit a characteristic
shape: a rapid rise in brightness followed by
a leisurely fall
\citep{Pres1964,Jame1986,Smit1995}.

Although RR Lyr stars have become important 
tools for the investigation of galactic structure,
they are not as well understood as one would wish
for such fundamental calibrators.
For example, we cannot compare rigorously our models of stellar
structure and their predictions for pulsation
to observations,
because we do not know precisely the mass of any 
RR Lyr star.
The reason is simple:
despite a few false alarms
\citep{Sosz2003,Prsa2008},
and one case -- TU UMa -- of what may be a very wide
binary containing an RR Lyr
\citep{Wade1999},
we have found no RR Lyr stars in eclipsing
binary systems which can be studied via photometric
and spectroscopic methods.
The discovery of even a few RR Lyr in eclipsing 
binary systems would provide a very valuable
check to our understanding of these stars
and improve our use of them as distance indicators.
On the other hand, if comprehensive searches reveal
that RR Lyr stars occur in binary systems at rates 
far below that of other, similar, stars,
such as the pulsating W Vir variables which {\it have}
been seen in eclipsing binary systems
\citep{Sosz2008},
we may deduce some features of the evolutionary sequence 
which leads to RR Lyr stars.


\section{Analysis of the MACHO photometry}

We begin with the collection of data described in
\cite{Kund2008},
which includes measurements of 3256 stars in the Galactic Bulge
made during the course of the 
MACHO project
\citep{Alco1997,Alco1999}.
The data are available freely from
the MACHO collaboration\footnote{
http://wwwmacho.anu.edu.au/},
but we could not find the detailed description
of their analysis
promised
by the reference
``Cook et al. (2007, in preparation).''
We therefore do not know the particular procedures
used to identify these stars as
RR0 (= RRab) Lyr variables,
nor to determine their fundamental periods.
All we have are the results of that analysis:
photometry of several thousand RR0 Lyr stars
in the $B_M$ (blue) and $R_M$ (red) passbands of 
the MACHO project.

\begin{figure}
  \plotone{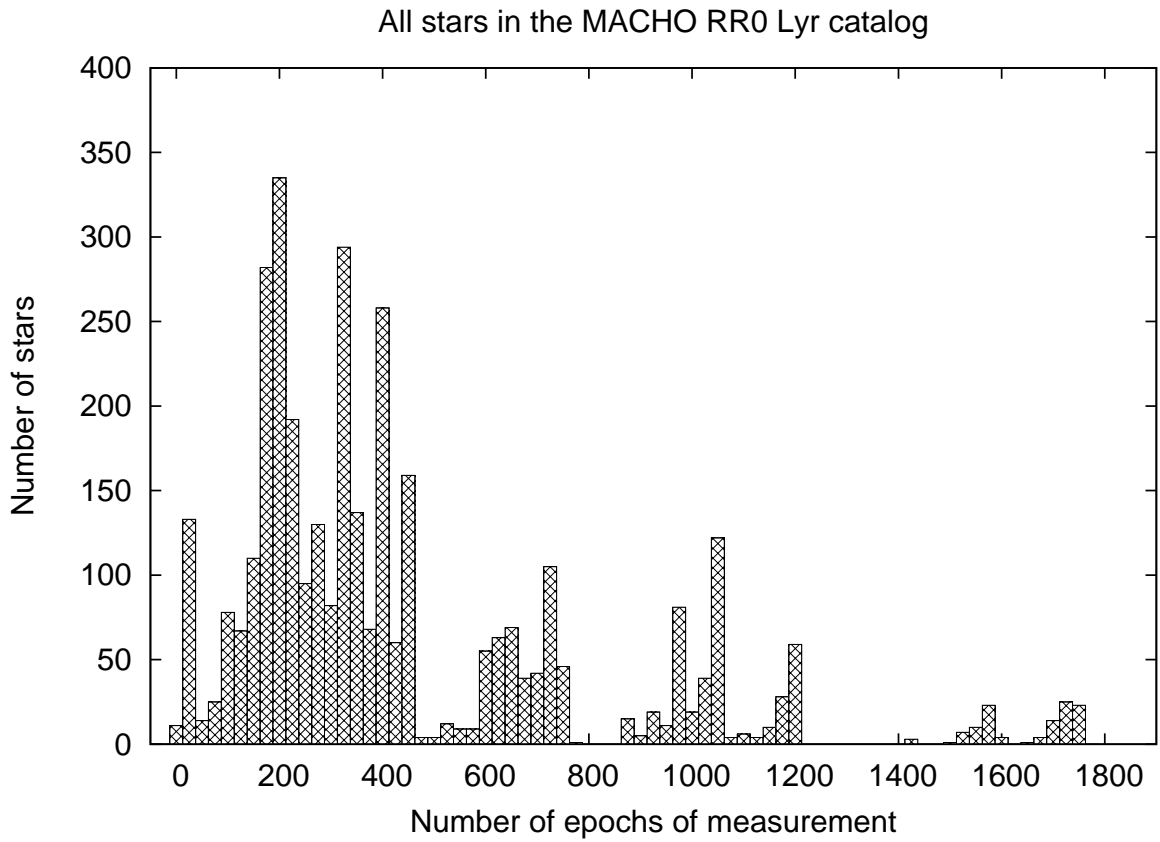}
  \caption{Number of epochs of measurement for stars in the
           MACHO Bulge RR0 Lyr catalog. \label{fig:numepochs} }
\end{figure}

Each star in this dataset is listed with 
its Right Ascension and Declination
and a MACHO identifier of the form
$NNN.xxxxx.sssss$,
in which $NNN$ identifies the field,
$xxxxx$ the tile,
and $sssss$ the star within that tile.
We will use this MACHO identifier as a label
for particular stars throughout this paper.
The data for each star consists of a fundamental period,
a mean $V-$band magnitude,
and a series
of measurements:
the Julian Date, red magnitude and estimated uncertainty,
blue magnitude and estimated uncertainty.
The mean $V-$band magnitudes fall largely in the range
$16 < m_{V} < 19$, but
the red and blue magnitudes listed for each measurement
are on an instrumental system and lie
between $-5$ and $-8$.
Figure \ref{fig:numepochs}
shows a histogram of the number of epochs of measurements of each star,
which is typically several hundred.

For the benefit of readers who may decide to use this database
for their own work,
let us mention that some caution is required.
Some measurements ought to be discarded:
for example, 
those marked with magnitude values of -99 or
magnitude uncertainties of 9.999.
We found that others are
so noisy that they provide no significant information.
Both the ``crowding'' and ``FWHM'' attributes associated
with each measurement can be used to identify data of
low significance. 
After examining the pattern of outliers in several 
test cases, 
we decided to discard any measurements in which the
``FWHM'' values for both the red and blue images
was larger than 6.5 pixels.

\begin{figure}
  \plotone{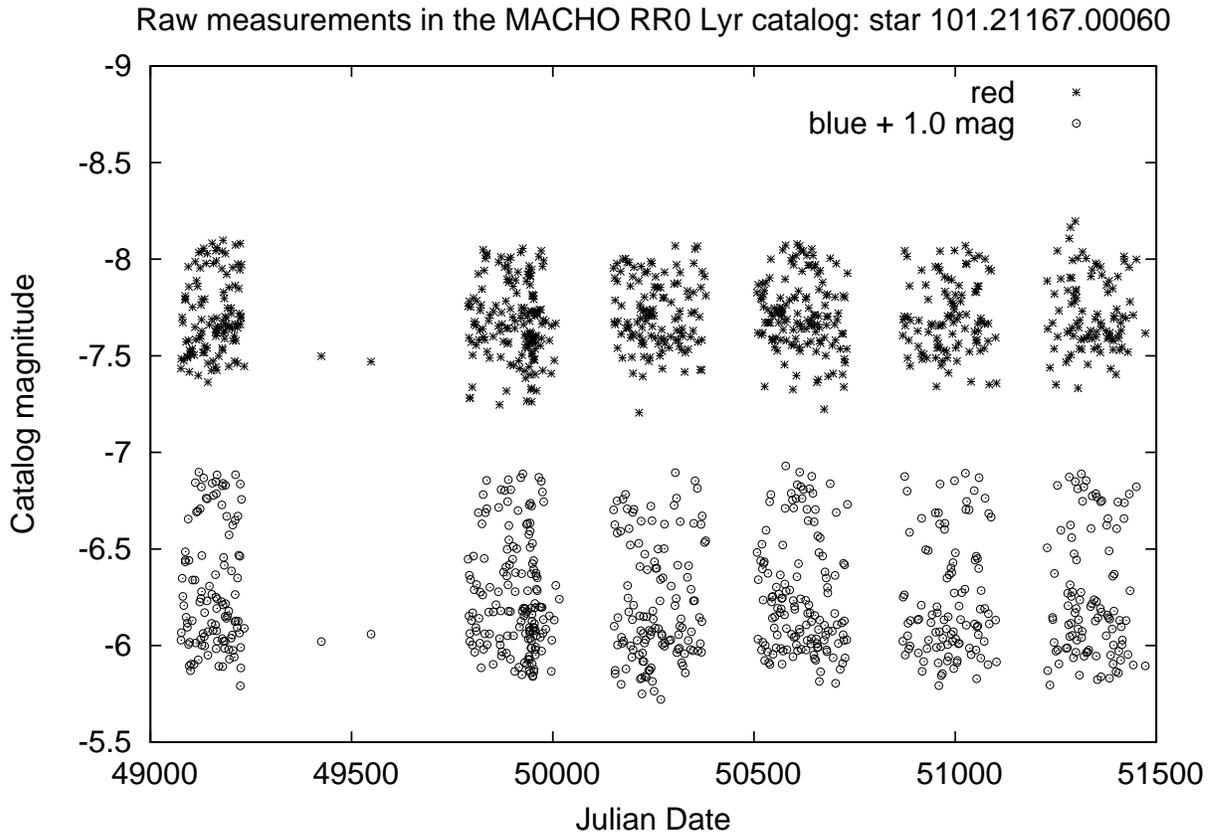}
  \caption{Photometry of one bright star. \label{fig:rawphotom} }
\end{figure}

This catalog includes measurements made over a span
of seven austral winters,
starting in April, 1993, and ending in October, 1999,
but most fields were not observed during all seven seasons.
Let us choose a single star as an example, and follow it 
through our analysis.
Star 101.21167.00060 is one of the brighter stars in the
catalog, with a mean $V-$band magnitude of 16.34.
A graph of its photometry,
Figure \ref{fig:rawphotom},
reveals that its field was not part of the regular
observing sequence during the second season.

\begin{figure}
  \plotone{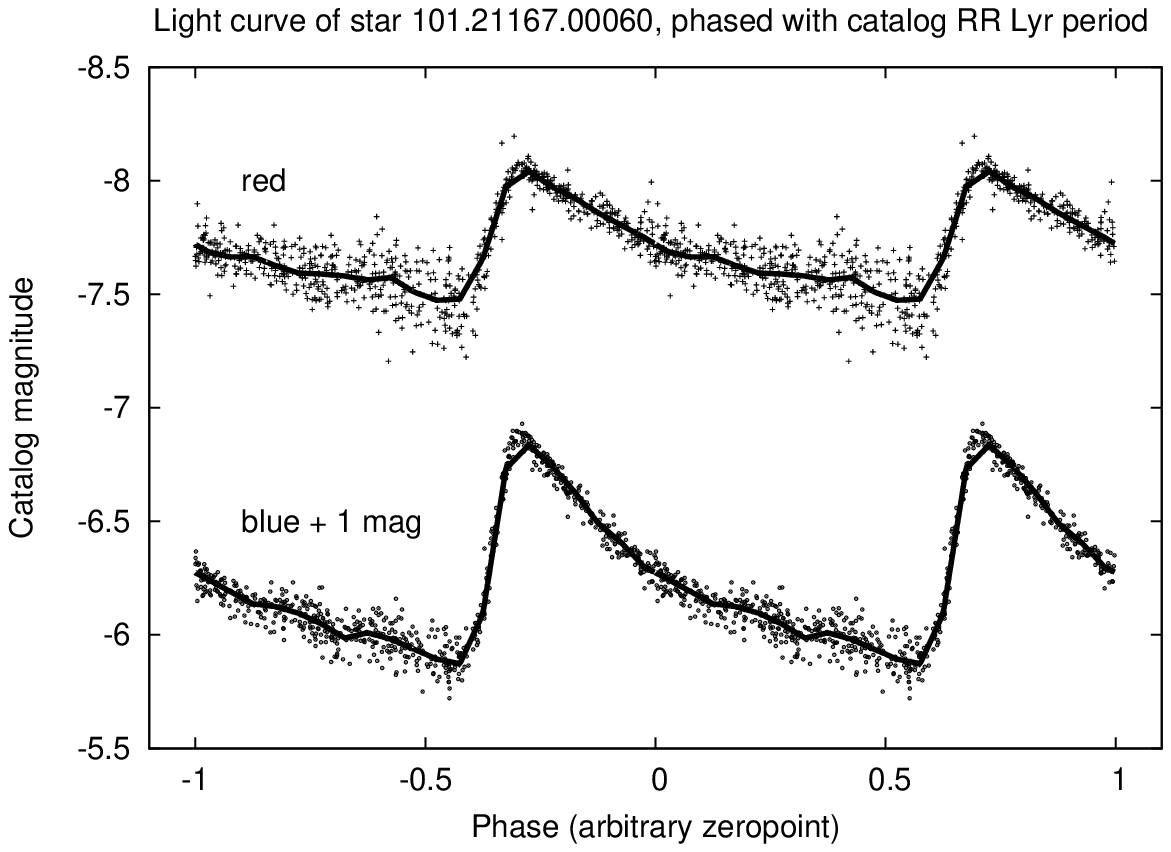}
  \caption{The model light curves for one bright star. 
           \label{fig:phasedrrlyr} }
\end{figure}

The goal of this work is to seek evidence of eclipses
in the light curves of RR Lyr stars.
It would be easier to find such evidence if the large variations
in light
due to the RR Lyr pulsations are removed.
Therefore, we created a model for the regular RR Lyr light curve
of each star and subtracted it from the measurements,
leaving the residuals for further consideration.
Our method to create the model for each star was simple:
we phased the data with the period given in the MACHO catalog,
divided the data into 20 bins of equal size in phase,
and computed the median magnitude within each bin.
We assigned this median magnitude to the phase in the middle
of its bin.
Finally, we interpolated linearly between these median
values to determine the magnitude at any phase.
Figure \ref{fig:phasedrrlyr},
shows the resulting models -- one for the red magnitudes,
one for the blue magnitudes -- for star
101.21167.00060.

Note that this simple approach has an obvious drawback:
it does not match the actual light curve well in places
where there is sharp change,
such as phase 0.7 in 
Figure \ref{fig:phasedrrlyr}.
However, it does provide a reasonable model for stars with
relatively few measurements, and it handles noisy data
very well.

\begin{figure}
  \plotone{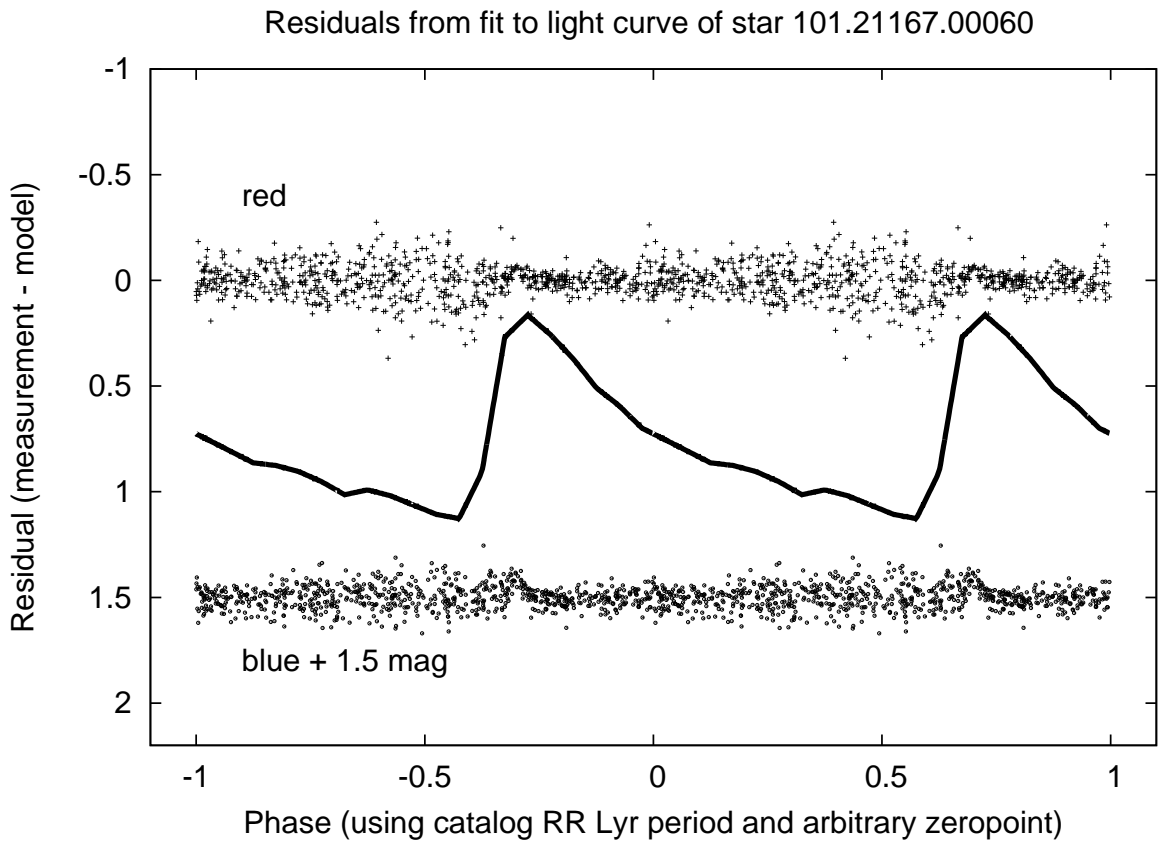}
  \caption{Residuals from the model light curves for one bright star;
           the blue model curve is shown, offset from the
           measurements for clarity.
           \label{fig:phasedresiduals} }
\end{figure}

After creating separate models for the red and blue measurements
of each star,
we subtracted the model from the data, 
leaving a set of residual magnitudes.
We show an example of these residuals,
phased with the RR Lyr period,
in 
Figure \ref{fig:phasedresiduals}.

\section{Identifying eclipsing binary candidates}

Having subtracted model RR Lyr light curves 
from the measurements of each star
in the red and blue passbands,
our next task was to search for patterns in 
the residuals which might indicate 
eclipses.
There are many approaches to this problem,
in general, but given the nature of our data --
very inhomogeneous sampling with large gaps and
often high noise levels --
and the uncertain nature of our expected signal --
which could range from sharp, narrow dips in light
to smooth, continuous variation --
we chose the ``string length'' method
\citep{Dwor1983,Bhat2010}.

Our implementation of this technique follows closely 
the description given in 
\citet{Bhat2010}.
We generated string lengths for periods between
0.10 and 100 days, using steps equally spaced in frequency
of size 0.0001 cycles per day. 
For each possible period, we computed a string length
separately for the red and blue measurements,
then added the two lengths to form an overall figure
of merit for that period.
We set thresholds for significance following the suggestions of
\citet{Dwor1983}
and ignored periods which exceeded these thresholds.
We saved the periods which yielded the 10 shortest
string lengths for further consideration.
In addition, we computed the string length for a
period of 9999 days; since this was much longer than
the actual span of observations, it yielded a ``phased''
light curve which was simply in chronological order.
Stars with very long periods of variation would show
a short string length for this artificial period.

\begin{figure}
  \plotone{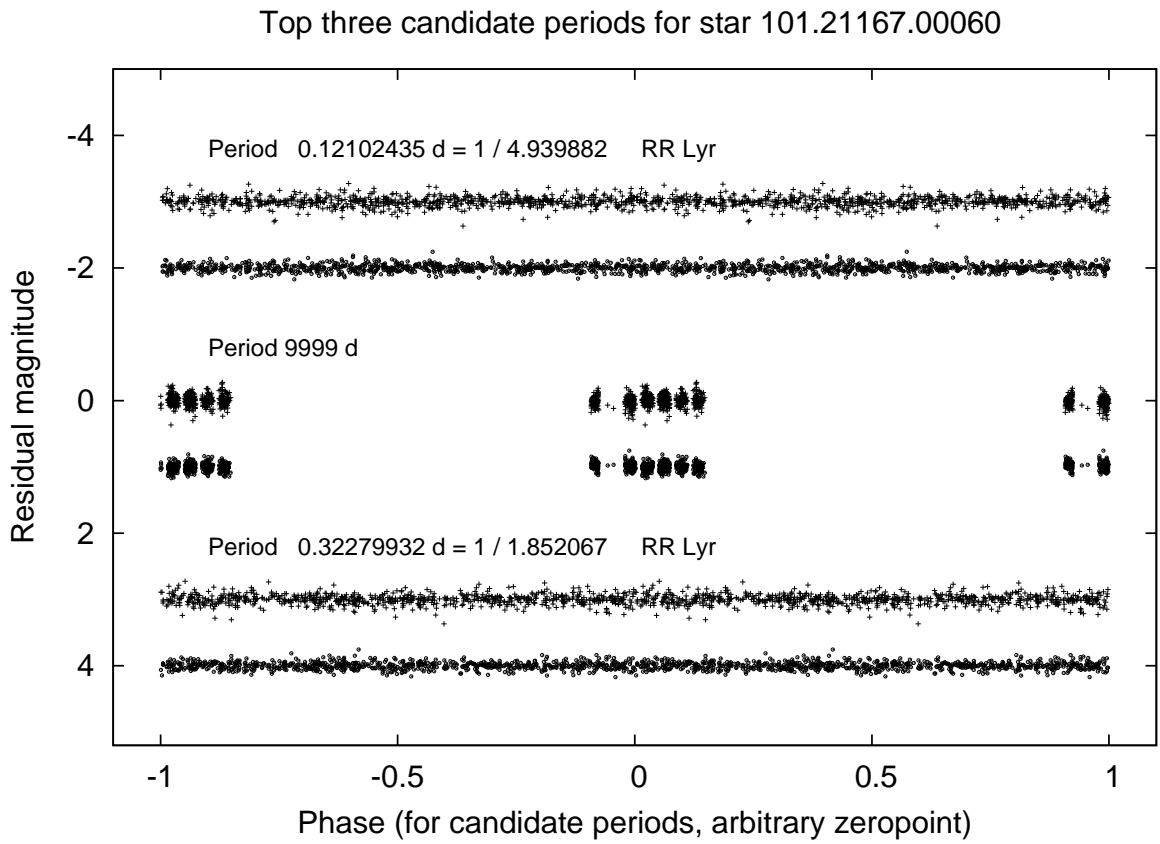}
  \caption{Light curves for the residuals of a bright star, phased with
           the best three candidate periods.
           \label{fig:phasedcandidates} }
\end{figure}

The next step was to examine the results for each star
visually.
We created a graphical representation of the star's light
phased with the best three periods,
as shown in 
Figure \ref{fig:phasedcandidates}.
In addition to the measurements, the graph
displayed the candidate periods, both in days and as a fraction
of the star's RR Lyr period.
A relatively quick view of this graph was sufficient 
to decide if any of the candidate periods yields
any significant signal,
and if the candidate periods are simply multiples or
fractions of the RR Lyr period.
The author examined graphs for all 3256 stars in
the catalog and noted those which deserved 
further consideration.

Stars which did show promising signs were subjected
to additional tests.
First, light curves were generated for all 10 of the
best candidate periods, in order to see which candidate
period looked most significant to the eye.
Second, 
we sometimes tried new periods, generating 
graphs and string lengths manually,
in order to yield a phased light curve with
2 maxima.
For example, if the best period $P$ created a
light curve with 5 maxima, we checked the
period $0.4P$.
When the results looked good,
we replaced the best automatically generated candidate
period with the manually chosen period.

\begin{figure}
  \plotone{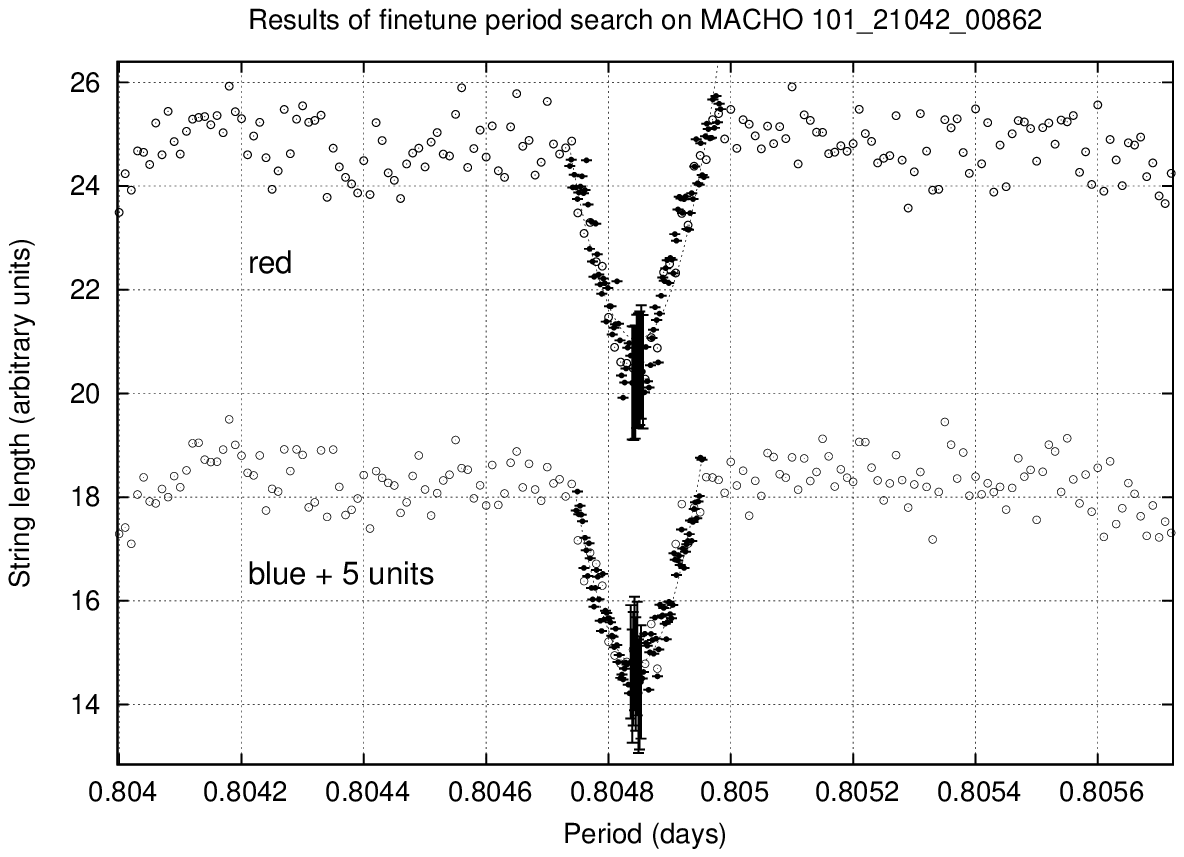}
  \caption{Determining the uncertainty in the best candidate
           period; the symbols with vertical errorbars
           span the uncertainty in period.
           \label{fig:finetune} }
\end{figure}

The final step was to improve the value of the
best candidate period, 
and to derive an estimate for the uncertainty in that period.
We set a small range around the best candidate period,
extending 0.05 days in both directions,
and divided the range into 10,000 pieces.
After computing string lengths for each piece,
we identified the range of periods which led to 
a local minimum in string length
(see Figure \ref{fig:finetune}).
Following the precepts of
\citet{Bels1983}
and
\citet{Fern1989},
we fit a parabola to this local minimum
and used the parameters of the fitted curve
to estimate both the period 
and the uncertainty in the period.

\begin{figure}
  \includegraphics[width=115mm,angle=-90]{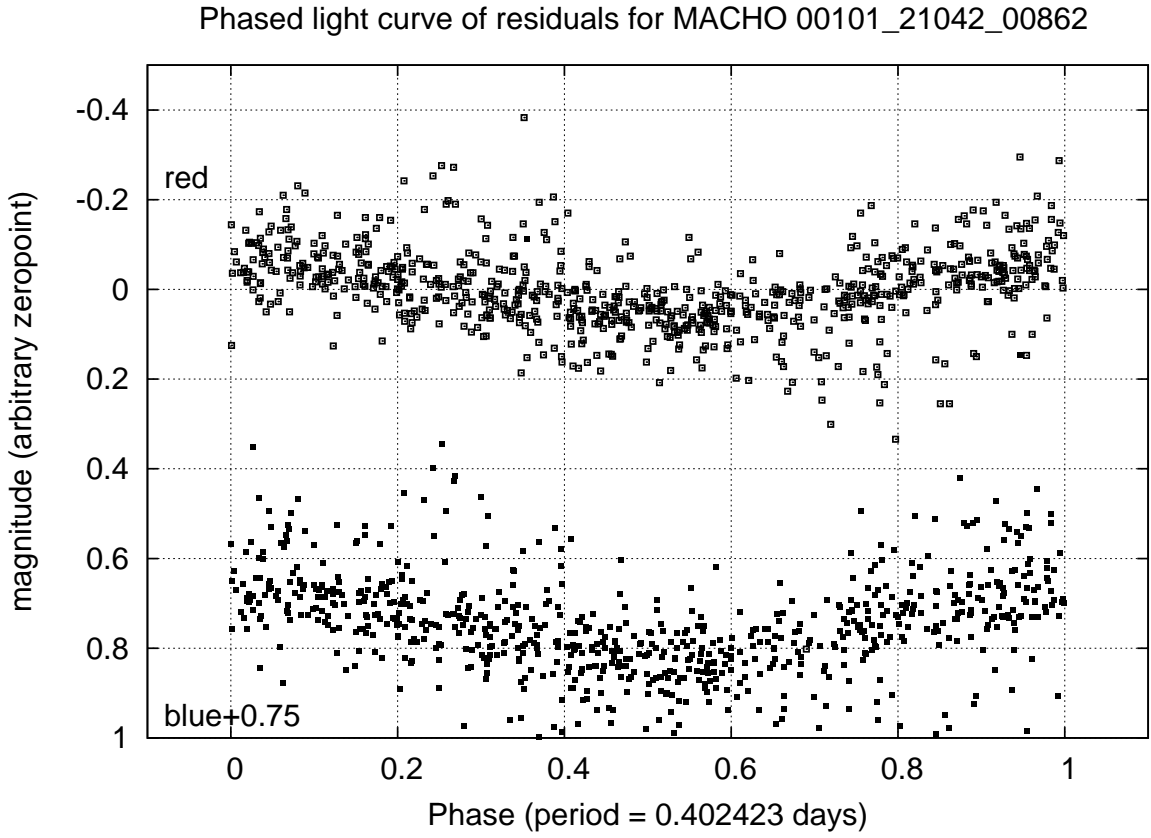}
  \caption{Phased light curve of star 101.21042.00862.
           \label{fig:cand1} }
\end{figure}

\begin{figure}
  \includegraphics[width=115mm,angle=-90]{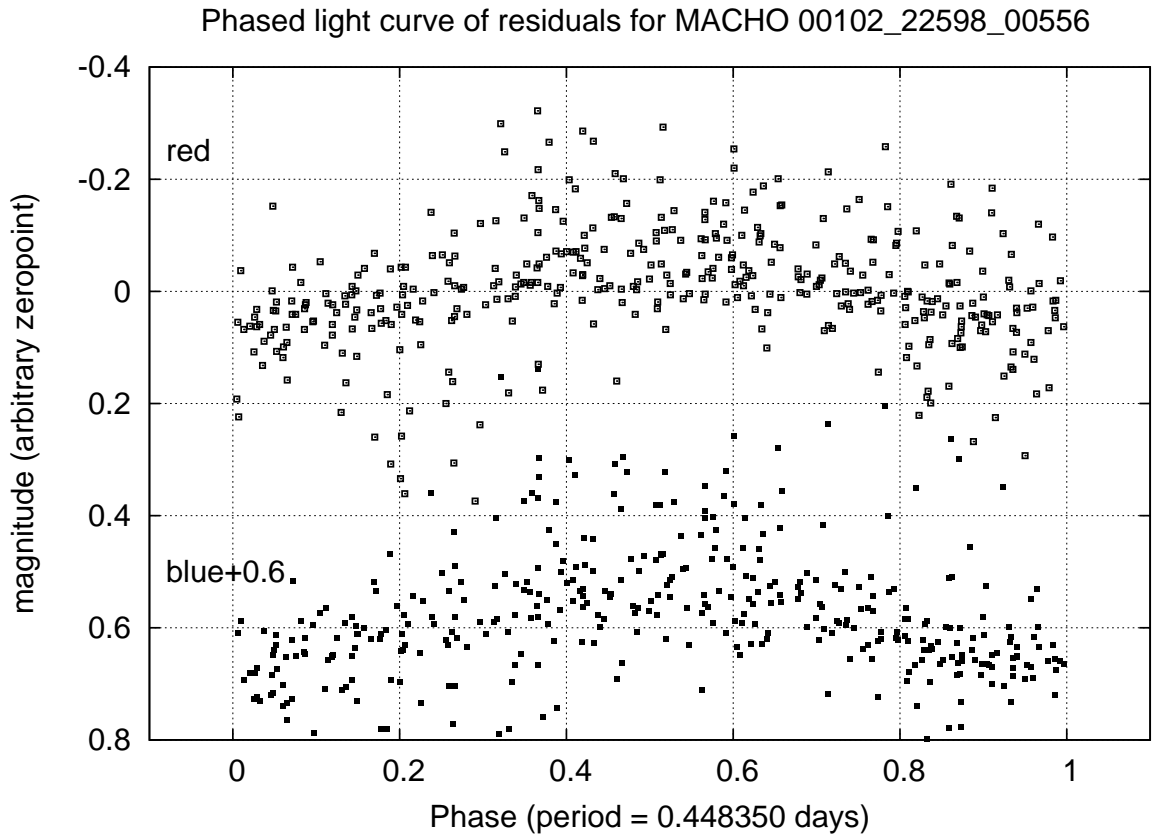}
  \caption{Phased light curve of star cand 102.22598.00556.
           \label{fig:cand2} }
\end{figure}

\begin{figure}
  \includegraphics[width=115mm,angle=-90]{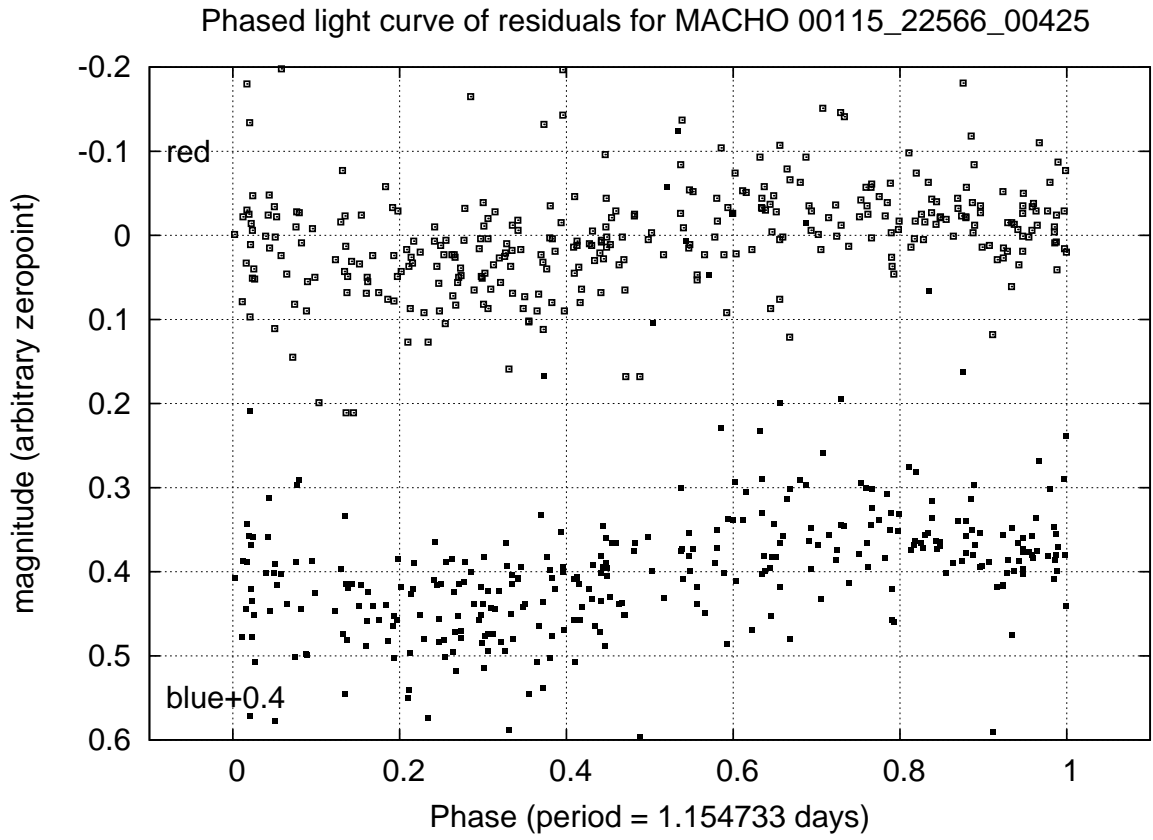}
  \caption{Phased light curve of star 105.22566.00425.
           \label{fig:cand3} }
\end{figure}

\begin{figure}
  \includegraphics[width=115mm,angle=-90]{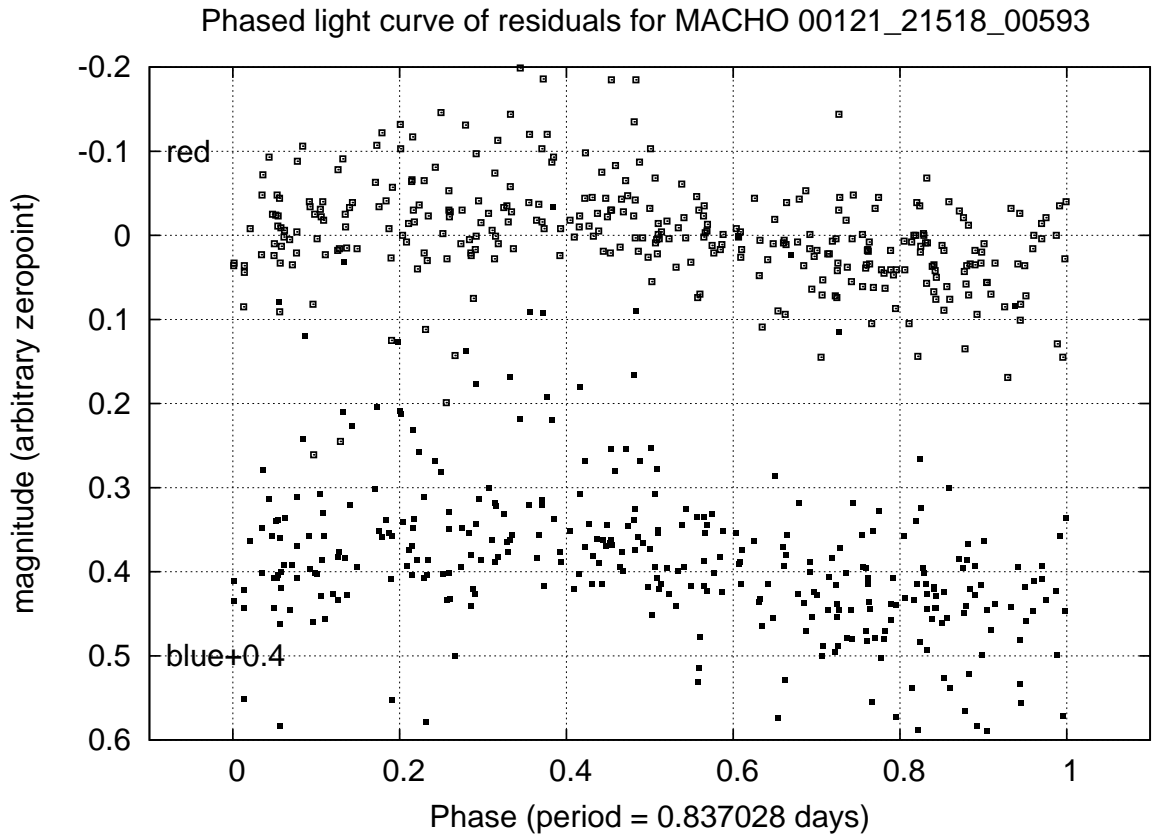}
  \caption{Phased light curve of star 121.21518.00593.
           \label{fig:cand4} }
\end{figure}

\begin{figure}
  \includegraphics[width=115mm,angle=-90]{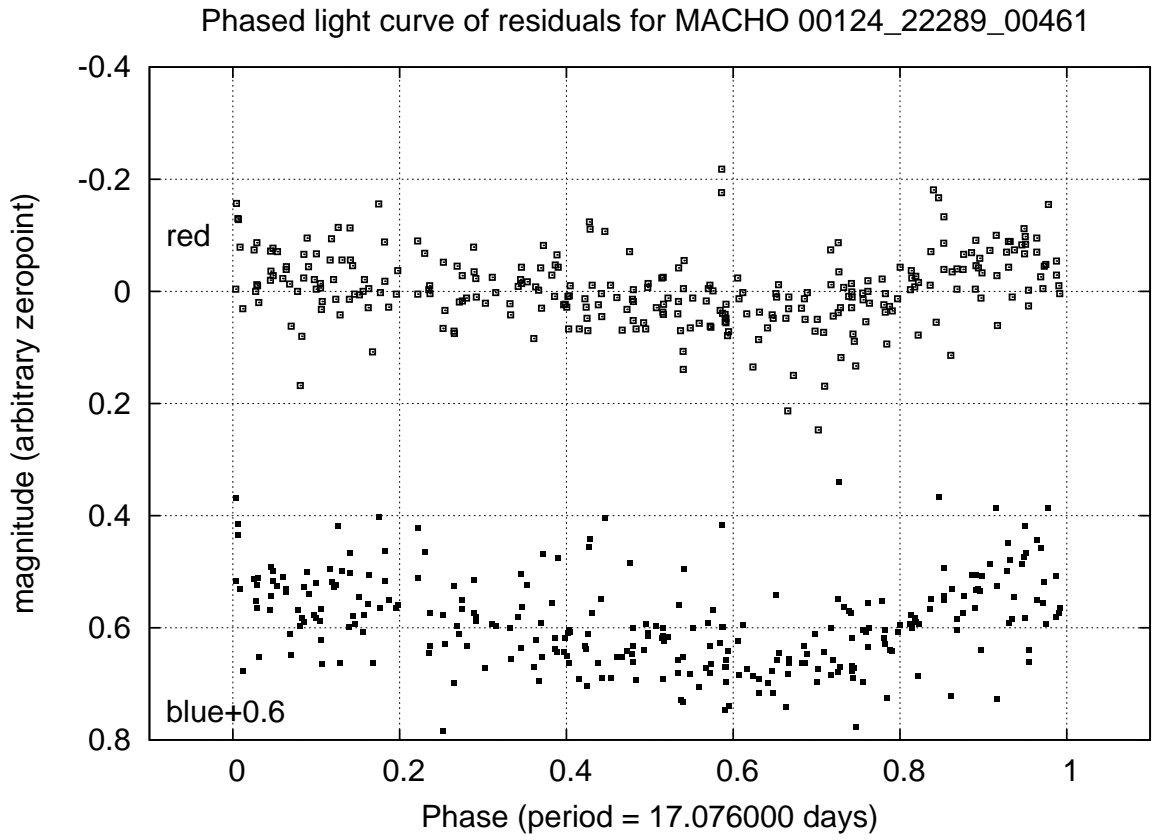}
  \caption{Phased light curve of star 124.22289.00461.
           \label{fig:cand5} }
\end{figure}

\begin{figure}
  \includegraphics[width=115mm,angle=-90]{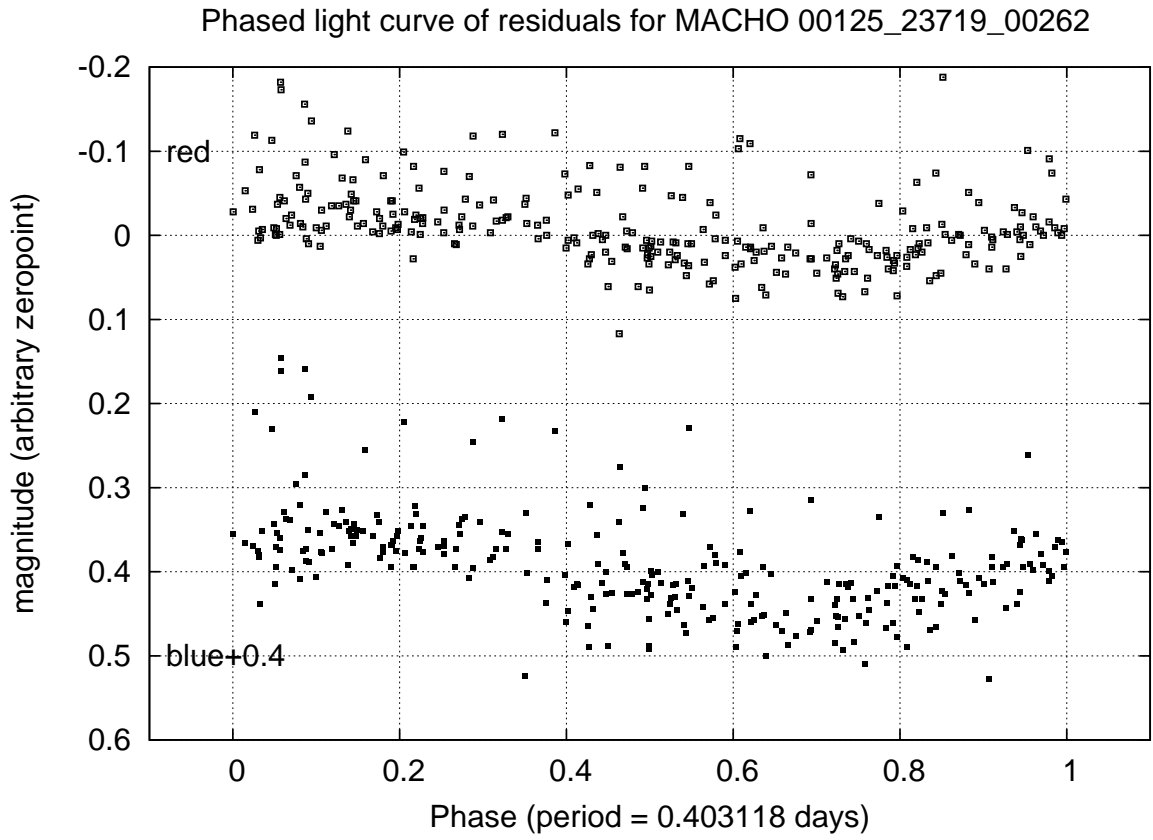}
  \caption{Phased light curve of star 125.23719.00262.
           \label{fig:cand6} }
\end{figure}

\begin{figure}
  \includegraphics[width=115mm,angle=-90]{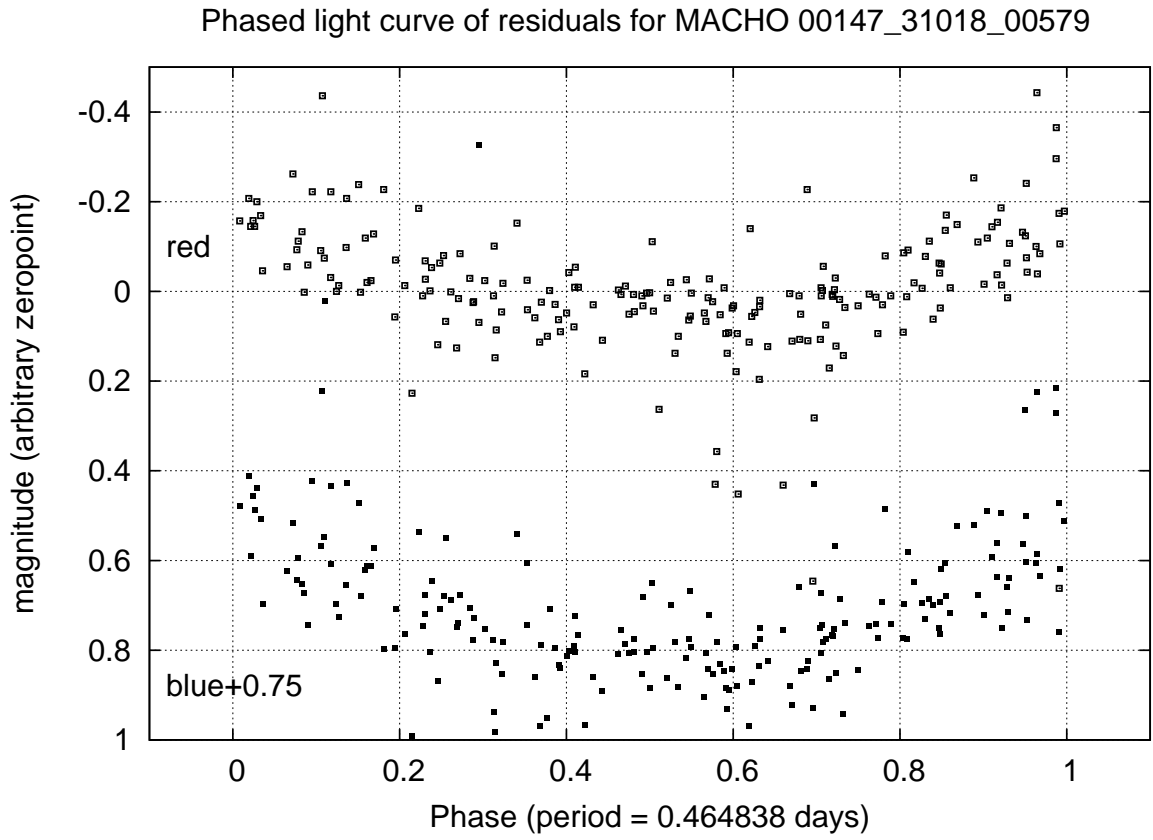}
  \caption{Phased light curve of star 147.31018.00579.
           \label{fig:cand7} }
\end{figure}

\begin{figure}
  \includegraphics[width=115mm,angle=-90]{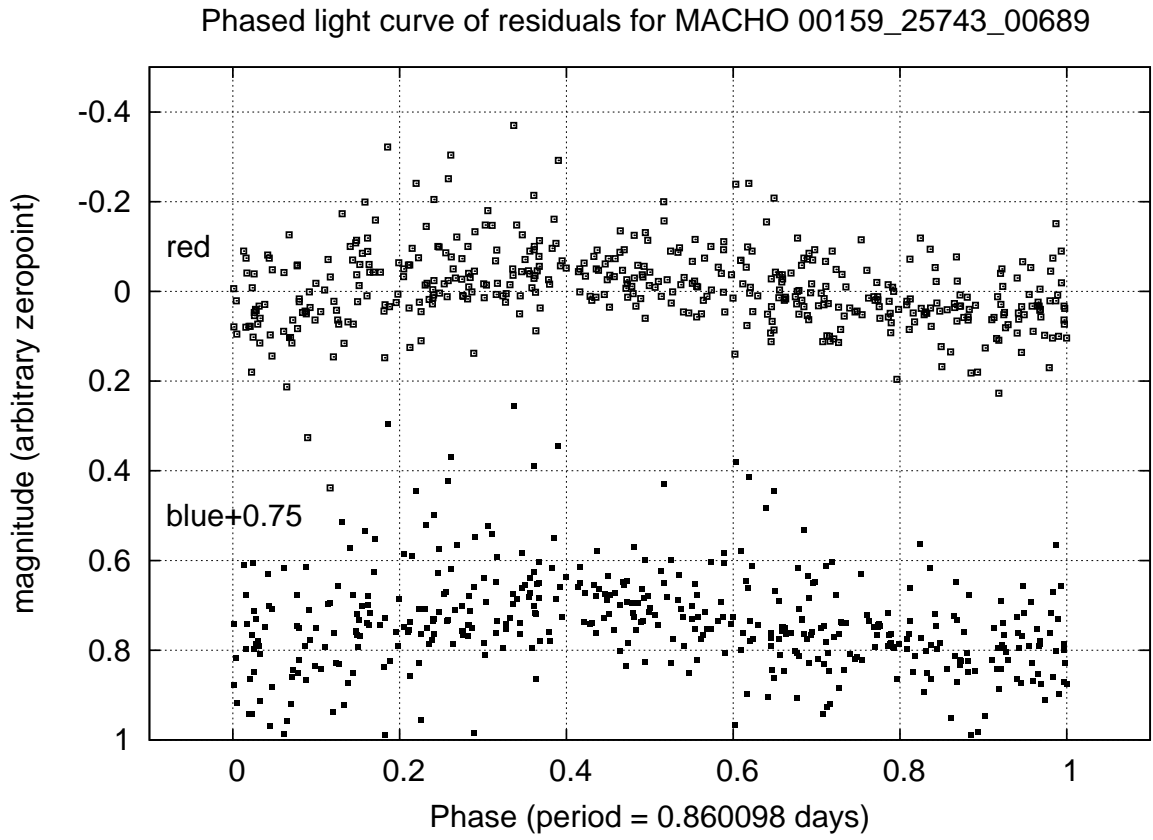}
  \caption{Phased light curve of star 159.25743.00689.
           \label{fig:cand8} }
\end{figure}

\begin{figure}
  \includegraphics[width=115mm,angle=-90]{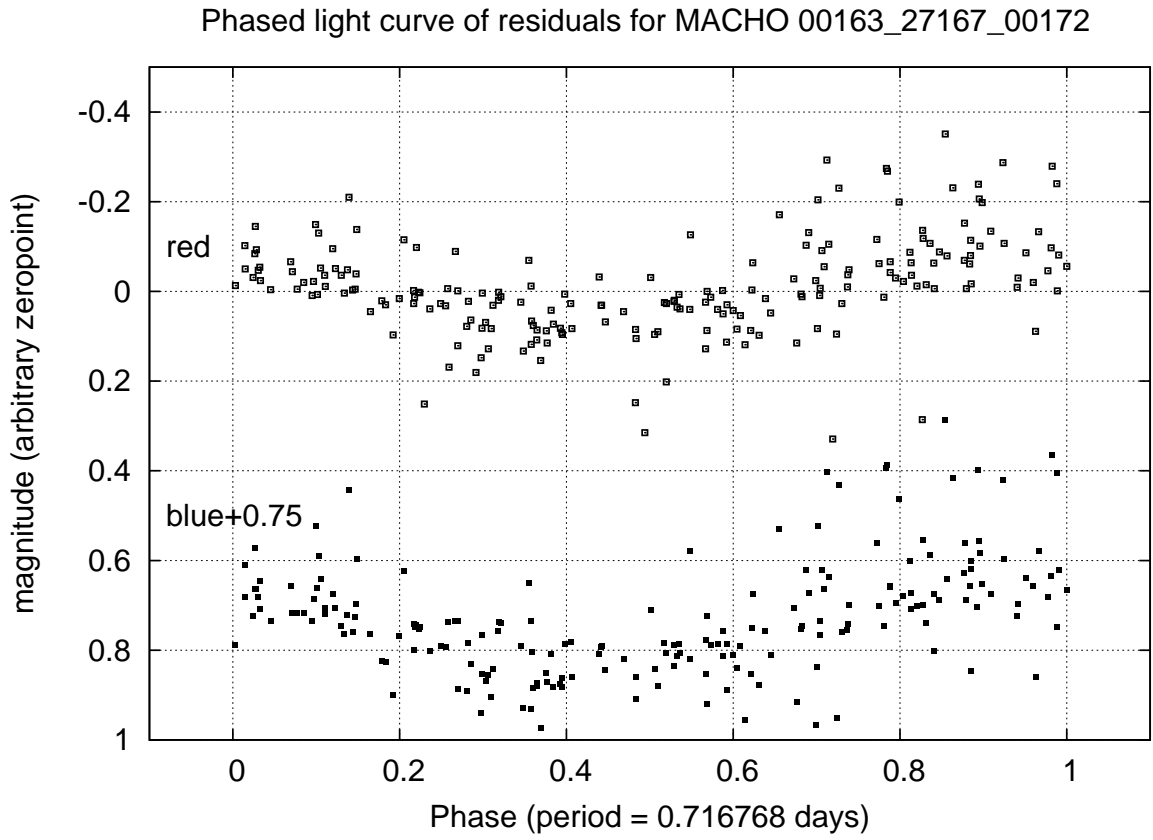}
  \caption{Phased light curve of star 163.27167.00172.
           \label{fig:cand9} }
\end{figure}

\begin{figure}
  \includegraphics[width=115mm,angle=-90]{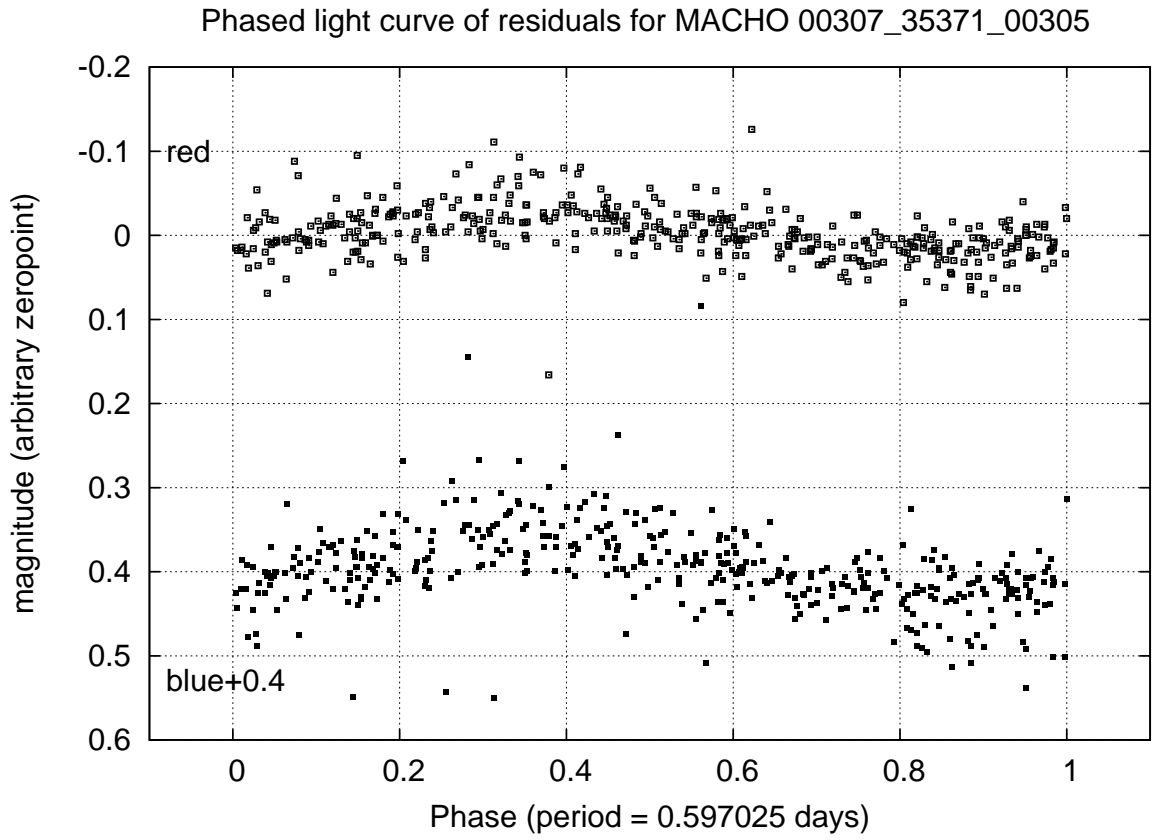}
  \caption{Phased light curve of star 307.35371.00305.
           \label{fig:cand10} }
\end{figure}

\begin{figure}
  \includegraphics[width=115mm,angle=-90]{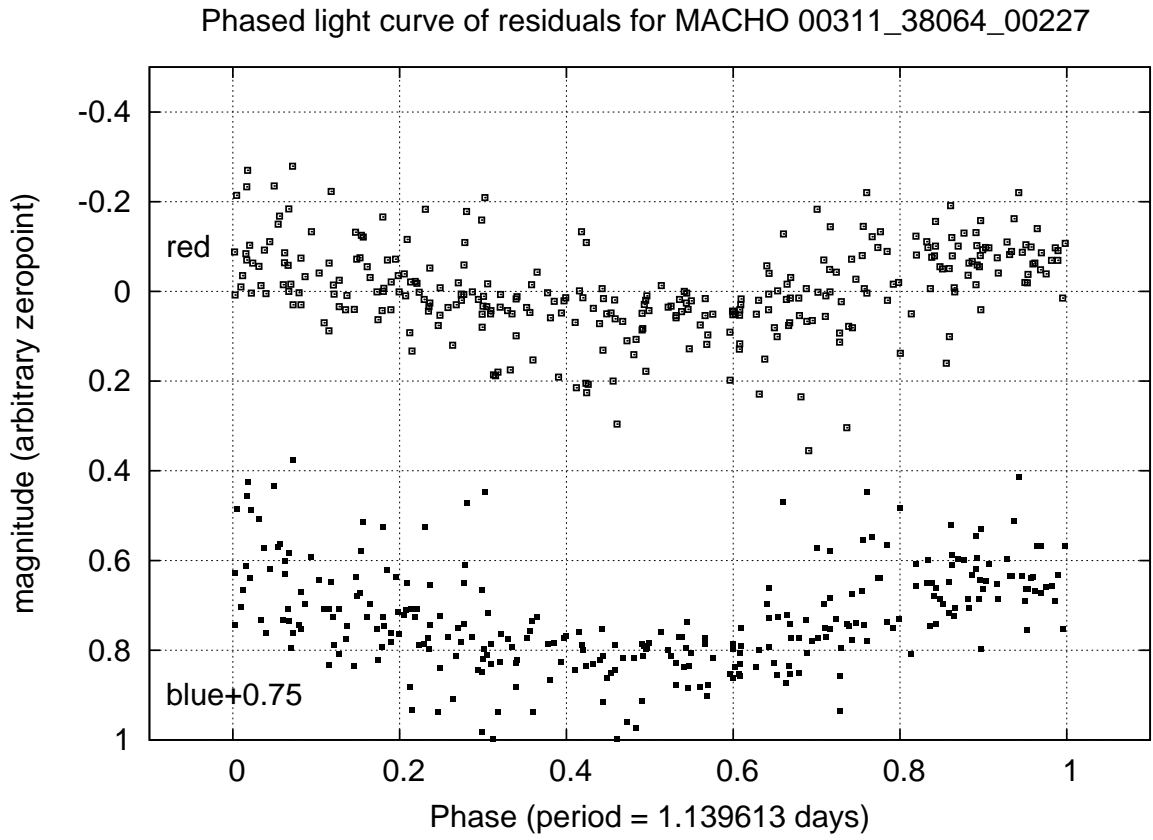}
  \caption{Phased light curve of star 311.38064.00227.
           \label{fig:cand11} }
\end{figure}

Based on a visual scan of the phased light curves
of the residuals for the best three candidate periods,
we selected 25 of the 3256 stars in the 
RR0 Lyr catalog for further analysis.
After additional checks,
we found that
11 
stars show clear, periodic patterns in their residuals.
We list these 11 stars,
which we shall call ``candidates'',
in Table
\ref{table:candidates}.
Values in the column labelled ``main period'' are taken from
the catalog of
\cite{Kund2008},
while those in the column labelled ``residual period''
were determined by this paper.
Phased light curves of each candidate are shown
in 
Figures
\ref{fig:cand1}
to
\ref{fig:cand11}.

\begin{center}
 \begin{table*}[ht]
 \caption{Candidates with periodic residuals}
 \label{table:candidates}
 {\small
 \hfill{}
 \begin{tabular}{ l l l r l l l l }
 \hline
 MACHO ID & 
 $\thinspace\thinspace$ RA$^a$ & 
 $\thinspace\thinspace$ Dec$^a$ &
        $V^{b} \thinspace$ &
        main period$^c$  &
        residual period$^c$  &
        amp$^d$  &
        notes$^e$ \\
 \hline

 101.21042.00862 & 18:05:18.60 & -27:16:14.5 & 17.55 & 0.413096 & $0.402423 \pm 0.000005$ &  0.13 & A \\
 102.22598.00556 & 18:08:54.96 & -27:31:49.4 & 17.36 & 0.444434 & $0.448350 \pm 0.000005$ &  0.13 & A \\
 115.22566.00425 & 18:09:09.00 & -29:40:30.4 & 16.83 & 0.530609 & $1.15473 \phantom{0} \pm 0.00004$  &  0.08 & D \\
 121.21518.00593 & 18:06:25.92 & -30:12:03.2 & 17.44 & 0.452635 & $0.83703 \phantom{0} \pm 0.00003$  &  0.07 & C \\
 124.22289.00461 & 18:08:15.72 & -30:50:30.5 & 17.72 & 0.498851 & $17.076 \phantom{00} \pm 0.002$    &  0.12 & \\
 125.23719.00262 & 18:11:45.96 & -30:50:13.2 & 16.47 & 0.684945 & $0.403118 \pm 0.000005$ &  0.10 & B \\
 147.31018.00579 & 18:28:43.68 & -29:31:55.6 & 18.45 & 0.528649 & $0.46484 \phantom{0} \pm  0.00001$  &  0.22 & \\
 159.25743.00689 & 18:16:18.12 & -25:52:22.4 & 17.56 & 0.459778 & $0.86010 \phantom{0} \pm  0.00003$  &  0.15 & B \\
 163.27167.00172 & 18:19:39.36 & -26:15:58.0 & 16.98 & 0.428079 & $0.71677 \phantom{0} \pm 0.00003$  &  0.16 & \\
 307.35371.00305 & 18:14:57.12 & -24:04:48.7 & 17.62 & 0.611346 & $0.59703 \phantom{0} \pm 0.00001$  &  0.07 & A \\
 311.38064.00227 & 18:19:33.60 & -23:46:16.0 & 16.86 & 0.528800 & $1.13961 \phantom{0} \pm 0.00005$  &  0.17 & C \\

 \hline
 \hline
 \multicolumn{8}{l}{\footnotesize $^{a}$ J2000 } \\ 
 \multicolumn{8}{l}{\footnotesize $^{b}$ mean magnitude, from Table 3 of Kunder et al.  (2008) } \\
 \multicolumn{8}{l}{\footnotesize $^c$ days  } \\ 
 \multicolumn{8}{l}{\footnotesize $^d$ magnitude of residual variation, peak-to-peak in blue band } \\ 
 \multicolumn{8}{l}{\footnotesize $^e$ A = residual period within 3\% of main period ; 
         B = residual period within 1\% of 1-day alias; C = within 2\% ; D = within 3\% 
     } \\
 \hline
 \end{tabular}}
 
 \end{table*}
\end{center}

\section{Discussion}

None of these candidates shows the sharp, narrow
dips of a detached system. 
We suggest two reasons to explain the absence of
such stars:
first, stars with dips of large amplitude may have
been excluded from the catalog of 
\citet{Kund2008},
if that catalog was constructed to contain
only stars with light curve shapes of an
isolated RRab variable.
Correspondence with the first author of 
\citet{Kund2008} 
suggests that this was not the case,
but we cannot dismiss it as a possibility.
Second, stars with dips of small amplitude may have
escaped our analysis due to the relatively low
signal-to-noise ratio of the individual measurements.
The dataset of
\citet{Kund2008}
includes an estimated uncertainty with each magnitude
measurement.  
We computed the mean value of the uncertainties for each star;
the median over 3565 stars of all those mean uncertainties
is 0.023 mag in the red band and 0.028 mag in the blue band.
A system of shallow eclipses, with an amplitude of only two or three times 
the typical uncertainty and involving only a small fraction of
the measurements, would not be found by our methods.
The smallest amplitude among our 11 candidates is
0.07 mag peak-to-peak, and that variation 
involves all the measurements
in the light curve, not just a few.

All the candidates have
gently undulating light curves,
rather than the sharp dips of 
a detached eclipsing system.
What might cause this sort of periodic
variation?
Among the possibilities are
artifacts from the subtraction of the model
light curve;
aliases of the RR Lyr frequency;
an additional frequency of oscillation in 
the RR Lyr star;
Blazhko variations in the RR Lyr star;
blended light from another variable star(s);
or the orbital motion of the RR Lyr star
around a close companion.
Let us examine these possibilities.

Our method for subtracting the main RR Lyr light curve
from each star's measurements was based on a simple
model made of linear segments.  If the model failed
to reproduce some features properly, the subtraction
would leave a signal with the same period as 
the RR Lyr pulsation.  Our 11 candidates include
3 stars for which the residual period is within 
3 percent of the main RR Lyr period;
we mark these 
in Table
\ref{table:candidates}
with a ``A'' in the ``Notes'' column.
It is possible that these candidates may be 
due to a low-amplitude version of the 
Blazhko effect, of which we say more below.

The majority of the measurements described in
\cite{Kund2008}
were collected on a nightly basis;
that is, each star was observed once per night,
and often at roughly the same time.
We therefore expect to see aliases 
of the true frequency $\omega_0$ in the
measurements with frequencies

\begin{equation}
 \omega_{\rm alias} = | \omega_0 \pm N \omega_{\rm sample} |
\end{equation}

\noindent
where $\omega_{\rm sample} = 1 \thinspace {\rm day}^{-1}$ 
and $N$ is some small integer.
We computed the alias frequencies for all candidates
using $N = 1$ and $2$, and compared them to the
frequencies of the residual variations.
We found two cases in which the residual frequencies
were within 1 percent of the main RR Lyr frequency
(marked ``B'' 
in Table 
\ref{table:candidates}),
two more cases within 2 percent
(marked ``C'')
and one more case within 3 percent
(marked ``D'').

Some RR Lyr stars are known to oscillate at
two frequencies;
these double-mode stars always have
a ratio of periods $P_1 / P_0 \simeq 0.746$ 
\citep{Neme1985,Szcz2007}.
None of our candidates have periods in their
residuals which yield this ratio with 
the periods listed in 
\citet{Kund2008}.
The lack of such double-mode pulsators
may not be unexpected, since
they appear to be very rare in the 
central regions of our Milky Way;
\cite{Mize2003} found only 3 such stars
among a sample of 1942 RRab and 771 RRc
stars observed near the center of the 
Milky Way in the OGLE-II database 
\citep{Udal1997}.
Their absence may also be a reflection of the
selection criteria used by 
\cite{Kund2008} 
to create the catalog of RR Lyr stars.

Some RR Lyr stars exhibit slow changes in the
shape and amplitude of their light curves,
with periods of tens to hundreds of days;
this is known as the Blazhko effect.
\cite{Mize2003} finds roughly 25 percent
of all RRab stars in a sample near the galactic
center show the Blazhko effect.
Could it be responsible for any of our
candidates?
MACHO 124.22289.00461, with a residual period of
just over 17 days, is the only candidate
for which this seems a possibility.
Unfortunately, since the main RR Lyr period
is almost exactly half a day, 
the measurements made during each observing season
cover only a small range in phase;
thus, we cannot see if the shape of the light
curve changes over this 17-day interval.
The three stars marked with an ``A'' in 
Table
\ref{table:candidates}
have residual periods which could be produced
by Blazhko-like periods of 15 to 50 days
beating against the main RR Lyr period;
however, we examined the phased light curves
of these stars over each season and see no
strong evidence of Blazhko variations.

The MACHO study area in the galactic bulge was, 
by design, chosen to have
a very high stellar density.  
If the density is high enough, we may expect
that blends of unrelated foreground or background
variable stars may cause periodic signals in 
the residuals of RR Lyr light curves.
Let us perform a very quick quantitative check
on this idea.
We examined one of the fields in this area,
number 102, counting all the stars in the
MACHO database (not just the variable ones)
as a function of apparent $V$-band magnitude.
Since the typical FWHM in these measurements
is $3{\rlap.}^{''}5$, a rough estimate of
the area of the seeing disk is about 10
square arcseconds.  
We find that, on average, there are
1.8 stars with $V \leq 20.1$ in each
seeing disk, 
and about 0.5 stars with $V \leq 18.1$.
The RR Lyr stars in the catalog of
\cite{Kund2008}
range roughly $16 \leq V \leq 19$.
so indeed a significant fraction of
all the RR Lyr stars in the catalog
must be contaminated by light from 
nearby stars at a level of 0.1 mag.
It may in fact be surprising that we find
so few objects with periodic variations
in their residuals;
however,
since we do not know the details of 
the process by which RR Lyr stars were
selected from the MACHO database,
we cannot comment further.

Could any of the candidates be due to the
effects of a binary companion of the RR Lyr star?
If two stars orbit each other with a separation 
which is only slightly larger than their combined
radii, 
their shapes may grow distorted enough that
they produce a gently undulating light curve,
even in the absence of eclipses.
An RR Lyr star of mass $0.7 M_{\odot}$
with a companion of equal mass
in a circular orbit of period
2 days
would have a separation of about 7 $R_{\odot}$.
The radius of a typical RR Lyr star varies
from about 4 $R_{\odot}$ to 6 $R_{\odot}$
\citep{Sodo2009},
leaving little room for a companion.
If an RR Lyr star did orbit a more compact companion
with a period in this range, it would surely be
greatly distorted, and so liable to vary in 
brightness as it moved in its orbit.
Whether an RR Lyr star would have stable pulsations
in such a close orbit is beyond the scope of
this paper.
Note that in this situation, the period listed
in Table
\ref{table:candidates}
would be {\it half} of the orbital period.

%
We conclude that our search through
a sample of 3256 RRab Lyr stars 
failed to find
any detached eclipsing binary systems,
and very likely failed to find eclipsing
systems of any sort with amplitudes of $\geq 0.07$ mag.
The implied disjunction between stars pulsing
in the fundamental RRab Lyr mode and stars 
in binary systems may provide clues to the
evolution of RRab Lyr stars.
Our simple technique for removing the ordinary 
variation of light in order to seek some signal
in the residuals would be well suited to the
more sinusoidal variations of RRc Lyr stars,
many of which can be found in the catalogs
of 
\cite{Sosz2009} 
and
\cite{Sosz2003}.

\acknowledgements
This paper utilizes public domain data obtained by the MACHO
Project, jointly funded by the US Department of Energy through the
University of California, Lawrence Livermore National Laboratory under
contract No. W-7405-Eng-48, by the National Science Foundation through
the Center for Particle Astrophysics of the University of California
under cooperative agreement AST-8809616, and by the Mount Stromlo and
Siding Spring Observatory, part of the Australian National University.
The author thanks Doug Welch for reviewing an early
version of this paper and providing many suggestions
which improved it.

\newpage

\end{document}